\documentclass{article}

\usepackage{arxiv}

\usepackage[utf8]{inputenc} 
\usepackage[T1]{fontenc}    
\usepackage{hyperref}       
\usepackage{url}            
\usepackage{booktabs}       
\usepackage{amsfonts}       
\usepackage{nicefrac}       
\usepackage{microtype}      
\usepackage{lipsum}
\usepackage{graphicx}
\usepackage{tabularx} 
\usepackage{amsmath} 
\usepackage{physics}
\usepackage{subcaption}
\usepackage{cite}

\title{Quantum Entanglement in Two-Photon Rabi Stark Model}

\author{
 Zakaria Boutakka \\
  Physics and Quantum Technology team, LPMC\\
  Ben M'sik Faculty of Sciences, Hassan II University\\
  Casablanca, Morocco \\
  \texttt{zakaria.boutakka-etu@etu.univh2c.ma} \\
   \And
 Zoubida Sakhi \\
  Physics and Quantum Technology team, LPMC\\
  Ben M'sik Faculty of Sciences, Hassan II University\\
  Casablanca, Morocco \\
  \texttt{zb.sakhi@gmail.com} \\
  \And
 Mohamed Bennai \\
  Physics and Quantum Technology team, LPMC\\
  Ben M'sik Faculty of Sciences, Hassan II University\\
  Casablanca, Morocco \\
  \texttt{mohamed.bennai@univh2c.ma} \\
}

\begin{document}
\maketitle
\begin{abstract}
In this is work, an investigation on the two-photon Rabi Stark model as a function of the coupling strength under the effect of different Stark coupling strength values is treated. Here, we numerically explore the spectral collapse of the \textit{2pRSM} as a function of the qubit-cavity field coupling strength to gain further physical insights. Also, the visualization of Wigner function in purpose to study the non-classicality in ground-state of the system. At the last, we measure the quantum entanglement via von Neumann Entropy for different ratios of the Stark coupling strength. This work deepens the understanding of the role played by the Stark coupling strength determining the quantum entanglement.
\end{abstract}


\section{Introduction}
The quantum Rabi model (\textit{QRM}) is a paradigmatic model in quantum optics \cite{Rabi1937}, which describes the basic interaction of a two-state system (atom or qubit) with a single quantized harmonic oscillator (electromagnetic field mode) through the following Hamiltonian

\begin{equation}
			H_{QRM} = \hbar \omega_{c} a^{\dagger}a + \frac{\hbar \omega_{0}}{2} \sigma_{z} + \hbar g \sigma_{x} (a^{\dagger}+a) 
\end{equation} 

where $\omega_{0}$ and $\omega_{c} $ are the qubit and oscillator frequencies, respectively, g is the interaction strength, $\{ \sigma_{z}, \sigma_{x} \}$ are two-state operators, and $a(a^{\dagger})$ is the annihilation (creation) operator for the oscillator. This model has been studied in various fields such as cavity quantum electrodynamics \cite{Raimond2001}, cold atoms \cite{Schneeweiss2018}, and trapped ion systems \cite{Leibfried2003}, its Hamiltonian predicts accurately many physical situations where an atom – artificial or real – is interacting with a confined cavity field.\\
Recently, the theoretical and experimental developments have been greatly inspired by \textit{QRM}. This model was motivated by the recent advances progress in materials science and nano fabrication have led to spectacular achievements in the single photon technologies, including the single photon generation in cold atoms \cite{Zhang2022, Farrera2016}, quantum dots \cite{Kiraz2004}, diamond color centers \cite{Janitz2020, Ju2021}, or superconducting circuits \cite{Wallraff2004}, and the single photon detection based on quantum entanglement \cite{Li2021, Slodička2013} or cross-phase modulation \cite{Matsuda2016}, etc…

Numerous models, derived from the \textit{QRM}, exhibit unique physical properties that distinguish them from the original Rabi model \cite{Zhu2020, Schiro2012, Buck1981}.  One such model is the quantum Rabi-Stark model, which integrates a freely adjustable nonlinear term. \\
The Rabi-Stark model has been extensively studied in theoretical research, uncovering a multitude of innovative features \cite{Chen2020, Maciejewski2015, Eckle2017, Xie 2017}. Significantly, this model exhibits a quantum phase transition when surpassing a specific critical value of the model’s parameter \cite{Chen2020}. Benefiting from development of quantum simulation technology, the so-called quantum Rabi-Stark model (\textit{QRSM}) has been realized by adding a nonlinear process $(H_{NL})$ to the \textit{QRM} \cite{Grimsmo2013, Grimsmo2014}, and the Hamiltonian is given by

\begin{equation}
			H_{QRSM} = H_{QRM}+ H_{NL} 
\end{equation} 

Grimsmo and Parkins developed an especially intriguing expansion of the model in 2013 \cite{Grimsmo2013}. These researchers wanted to know if it was possible to achieve the \textit{QRSM} using a single atom linked to a high-finesse optical cavity mode. They devised a technique in which two hyperfine ground states of a multilevel atom mimic an effective two-level system, with resonant Raman transitions between the two states driven by a cavity field and two auxiliary laser fields. Importantly, this technique enables the realization of the \textit{QRM} in which coupling constants and effective frequencies can be freely and independently changed, allowing for a systematic investigation of the ultrastrong and deep-strong coupling regimes \cite{Eckle2017}.

The Grimsmo-Parkins technique, on the other hand, necessitates the inclusion of a new term to the quantum Rabi Hamiltonian, a nonlinear coupling term between the two-level system and the quantum oscillator, for generic values of the model's parameters. A coupling term of this type has been studied in the quantum optics literature as the dynamical Stark shift, which is a quantum variant of the Bloch-Siegert shift \cite{Bloch1940}. As a result, the quantum Rabi model supplemented with a nonlinear factor of the type outlined by Grimsmo and Parkins will be referred to as the \textit{QRSM}.

However, the corresponding nonlinear coupling strength in the conventional dynamical Stark shift is defined by the parameters of the underlying quantum Rabi model. The Stark coupling can also be modified freely and independently in the approach presented by Grimsmo and Parkins.\\ 
The \textit{QRSM} may undergo a superradiant transition in the deep strong coupling phase of the Rabi coupling, according to Grimsmo and Parkins \cite{Grimsmo2014}, when the Stark coupling strength equals the frequency of the cavity mode. As a result, the additional nonlinear part in the Hamiltonian, the Stark term, may give rise to novel physics. It will thus be critical to thoroughly explore the spectrum features of the \textit{QRSM}.

On the other hand, there has also long been interest in the alternative nonlinear coupling, which takes the form of a qubit and two photons \cite{Felicetti2015, Duan2016, Yan2022}. Unlike to its one-photon counterpart, the two-photon coupling can induce some unusual features, like the collapse of spectra. This so-called two-photon \textit{QRM} (\textit{2pQRM}) \cite{Hammani2024} becomes a hot topic recently \cite{Felicetti2018, Cong2019, Casanova2018, Maldonado2019, Maciejewski2019, Dodonov2019, Villas2019} .\\

The basic idea of our work is we add the Stark coupling term to the \textit{2pQRM}, which can be named two-photon RSM (\textit{2pRSM}) \cite{Li2020, Yan2024}. The idea evoked in this work focuses on the induction of a Stark coupling in the expression of \textit{2pQRM} whose goal is the realizations of ground-state entanglement. In this frame, by using the quantum simulation software QuTiP (Quantum Toolbox in Python) which is an open source software developed by Johansson et al. \cite{qutip}, we can numerically explore the spectral collapse of the \textit{2pRSM} as a function of the qubit-cavity field coupling strength, as well as the visualization of Wigner function in purpose to study the nonclassicality in ground-state of the system, besides a measurement of the quantum entanglement via Von Neunmann Entropy for different ratios of the Stark coupling strength.

The structure of this paper is organized as follows: In Sec. \ref{model}, is dedicated to introduce in detail the \textit{2pRSM}. Then in Sec. \ref{results} we explore the numerical simulation and discussion of results. Finally, we present the main conclusions of this work in section \ref{conclusion}.

\section{Model}
\label{model}

\subsection{Rabi-Stark model}
As we have expounded in the introduction, the quantum Rabi model describes the interaction between light and matter, next to the Jaynes–Cummings model, in the simplest possible way and is used as a basic model in many fields of physics.\\
For the \textit{QRM} with tunable parameters, it has been realized in quantum simulations based on Raman transitions in an optical cavity QED settings \cite{Monroe1995}. In this proposed scheme, a new nonlinear coupling item between the atom and cavity $H_{NL} = \gamma \sigma_{z} a^{\dagger}a$ is added to the linear dipole coupling, where the coupling strength $\gamma$ is determined by the dispersive energy shift. 
This generalized model was first proposed by Grimsmo and Parkins, which can be called quantum Rabi-Stark model because the new added item is associated with the dynamical Stark shift discussed in the quantum optics. Its Hamiltonian is expressed as follows \cite{Xie2019-1}:

\begin{equation}
		\begin{aligned}
			H_{QRSM} &= H_{QRM} + H_{NL} \\
			&= \hbar \omega_{c} a^{\dagger}a + \frac{\hbar \omega_{0}}{2} \sigma_{z} + \hbar g \sigma_{x} (a^{\dagger}+a) + \gamma \sigma_{z} a^{\dagger}a
		\end{aligned}
\end{equation} 

where an additional term, $\gamma \sigma_{z} a^{\dagger}a$, appears compared to the original quantum Rabi Hamiltonian $H_{Rabi}$. This additional term models a nonlinear coupling between the two–level atom and the single–mode cavity oscillator. In the introduction, we gave an argument for naming this Hamiltonian and the corresponding model the quantum Rabi–Stark Hamiltonian and model, respectively, with the coupling constant $g$, the Stark coupling $\gamma$\\

\subsection{Two-Photon Rabi Stark model}
The Two-Photon Rabi Stark model (\textit{2pRSM}) should be very interesting at the present stage as a significant generalization of \textit{QRSM}. In the \textit{2pRSM}, the interaction term is quadratic in the annihilation and creation bosonic operators. 
The Hamiltonian of the \textit{2pRSM} is given by ($\hbar = 1$) :

\begin{equation}
			H_{2pRSM} = - ( \frac{\hbar \omega_{0}}{2}+ \gamma a^{\dagger}a) \sigma_{x}  + \hbar \omega_{c} a^{\dagger}a + \hbar g [(a^{\dagger})^{2}+a^{2}] \sigma_{z}
            \label{eq 2pRSM}
\end{equation} 

This model holds significant interest due to its relevance to the study of the interaction between a two-level quantum system and the electromagnetic radiation field, particularly in the context of the two-photon transitions.

Furthermore, such a model could potentially be realized within solid-state devices, including trapped ion systems and circuit QED. In fact, recent demonstrations have successfully realized the implementation of the one-photon RSM utilizing a single trapped ion \cite{Cong2020}. 

Within this framework, external laser fields are employed to facilitate an interaction between an electronic transition and the motional degree of freedom, thereby the Stark term is generated. Given that the two-photon quantum Rabi model (\textit{2pQRM}) has been successfully achieved in trapped ion systems \cite{Felicetti2015}, there exists a viable pathway for the experimental simulation of the \textit{2pRSM} through laser driving within the same experimental apparatus.

In this work we are interested in nonlinear coupling in the form of the qubit and two photons where the special feature spectral collapse can be driven by the two-photon coupling, as well as changes in this coupling and how it affects the system dynamics and the achievement of the quantum entanglement.

\section{Results \& Discussion}
\label{results}

In this section, we present the main findings derived from the numerical simulations, illustrating the enhancement of the performance metrics of our model. Initially, it is pertinent to mention that the computational software employed for our simulations is QuTiP, which offers readily available definitions of standard quantum states and operators, alongside a plethora of dynamic solvers and visualization tools. In this context, by exploiting the Hamiltonian \ref{eq 2pRSM}, we are able to numerically investigate the eigenenergy spectrum of the proposed system and other pertinent physical parameters, such as the nonclassicality in the system, in relation to the critical coupling strength.

\subsection{The Spectrum}
Now, we present our numerical results for the energy levels spectrum of the Two Photon Rabi–Stark model as function of the coupling strength $g$

 \begin{figure}[ht]
		\centering
        \begin{subfigure}{0.4\textwidth}
		\includegraphics[width=1.2\linewidth]{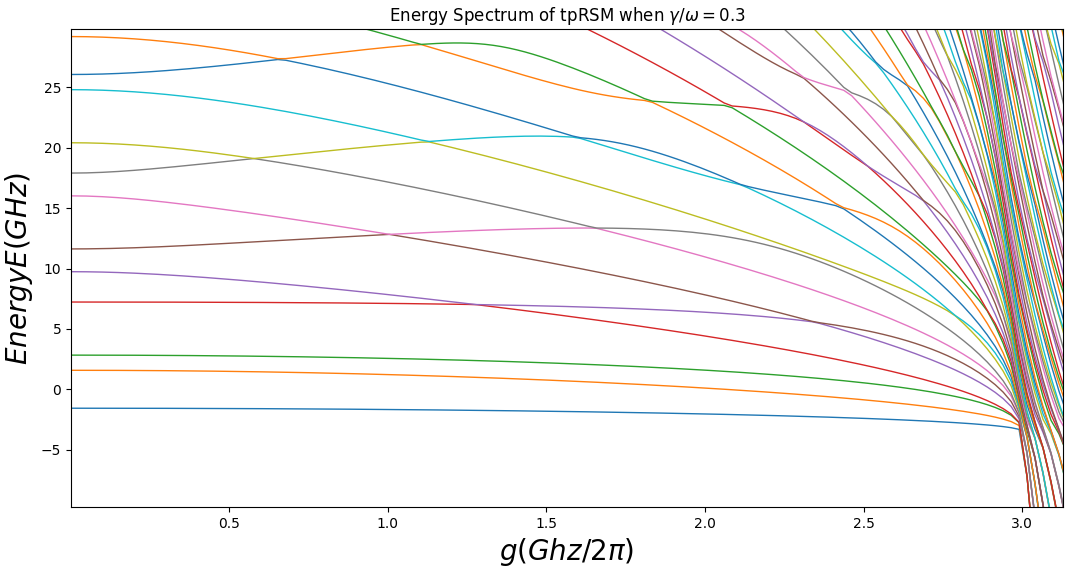}  
	\caption{}		
        \label{E0.3}
        \end{subfigure}
        \hfill
		\begin{subfigure}{0.4\textwidth}
		\includegraphics[width=1.2\linewidth]{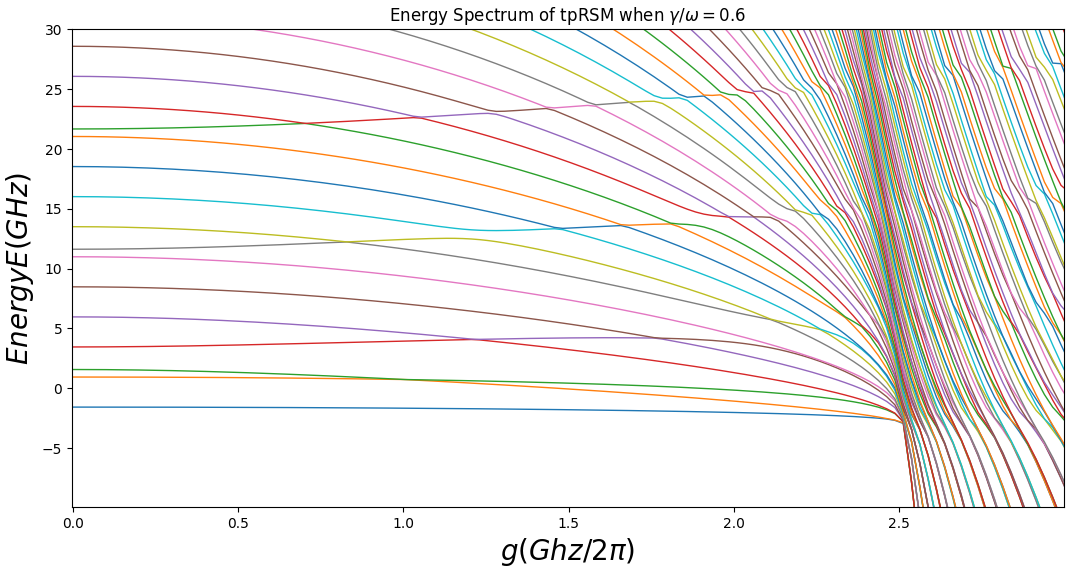}
        \caption{}
		\label{E0.6}
        \end{subfigure}
        \hfill
		\begin{subfigure}{0.4\textwidth}
		\includegraphics[width=1.2\linewidth]{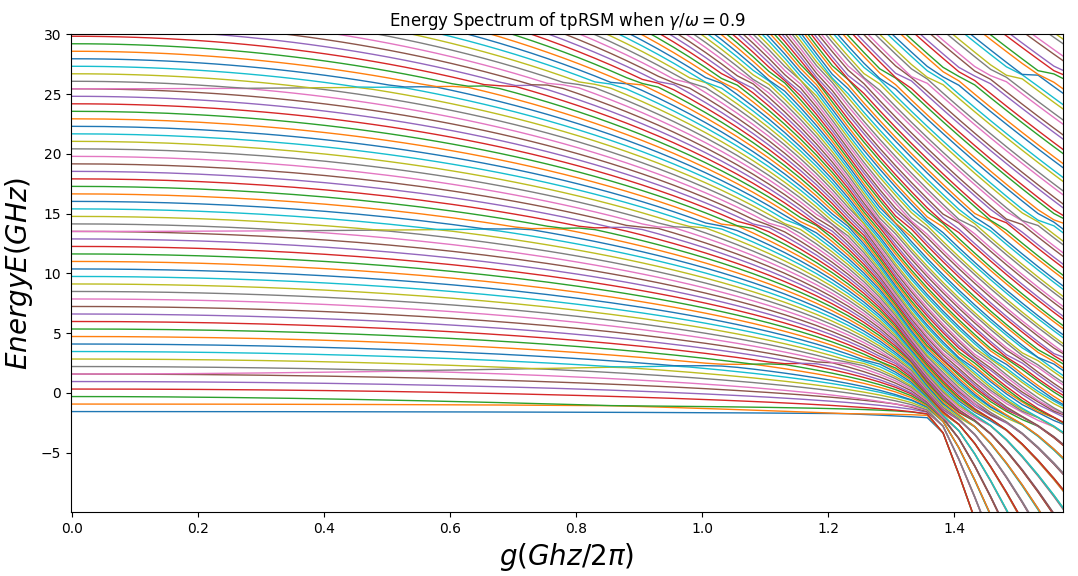}
        \caption{}
		\label{E0.9}
        \end{subfigure}
  
    \caption{The eigenspectrum obtained from the simulation of the Hamiltonian in Eq. \ref{eq 2pRSM} as function of the coupling strength  $g$ for different values of the Stark coupling strength $\frac{\gamma}{\omega}$ : a $\frac{\gamma}{\omega}= 0.3$,  b $\frac{\gamma}{\omega}= 0.6$ and c $\frac{\gamma}{\omega}= 0.9$, with number of cavity fock states N = 200.}
    \label{E 2pRSM}
\end{figure}

The Fig \ref{E 2pRSM} illustrate the energy spectrum of the two-photon Rabi-Stark model (\textit{2pRSM}) as a function of the coupling strength $g$ for different values of the Stark coupling strength $\frac{\gamma}{\omega}$. Observations from Fig. \ref{E0.3}-\ref{E0.9} indicate that as $g$ increases, the energy levels commence to exhibit clustering behavior and ultimately converge into a narrow band at critical coupling strength values $g_{c}$, which is indicative of the spectral collapse phenomenon.
In the $\frac{\gamma}{\omega}= 0.3$ case, energy levels are initially dispersed but they but start clustering around $g \approx 2.5$. As Fig. \ref{E0.3} illustrates, a considerable spectral collapse is seen near $g = 3$. 
As demonstrated in Fig. \ref{E0.6}, increasing $\frac{\gamma}{\omega}$ to 0.6 causes a more noticeable collapse at lower $g_{c}$ value in contrast to $\frac{\gamma}{\omega}= 0.3$. 
The energy levels start to cluster about $g \approx 1.2$ for $\frac{\gamma}{\omega}= 0.9$, at critical value $g_{c} = 1.4$ a significant collapse occurs compared to the previous cases (see Fig \ref{E0.9}). 

These findings elucidate that an elevation in the Stark coupling strength results a reduction of the threshold $g$ for spectral collapse, thereby indicating that augmented Stark interaction leads to pronounced aggregation of energy levels. This phenomenon is consistent with theoretical forecasts \cite{Felicetti2018, Peng2019, Lo2020} and demonstrates the important role of Stark coupling on the \textit{2pRSM} dynamics, which is essential for for applications requiring the control or avoidance of spectral collapse.

\subsection{Dynamics of the \textit{2pRSM}}
The Wigner function \cite{Wigner1932, Walther1978} is a quasi-probability distribution that is used to visualize the states of a harmonic oscillator with a particular application to the modes of light in quantum optics \cite{Leonhardt1997} and provides insight into the quantum state's phase space distribution. It is particularly useful for identifying non-classical properties like quantum interference and entanglement.

\begin{figure}  
        \centering
		\begin{subfigure}{0.4\textwidth}
			\includegraphics[width=1.2\linewidth]{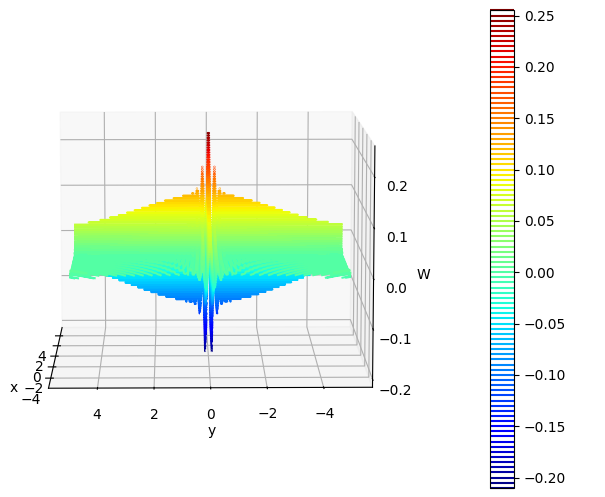}  
			\caption{}
                \label{}      
		\end{subfigure}
               \hfill
        \begin{subfigure}{0.4\textwidth}
			\includegraphics[width=1.2\linewidth]{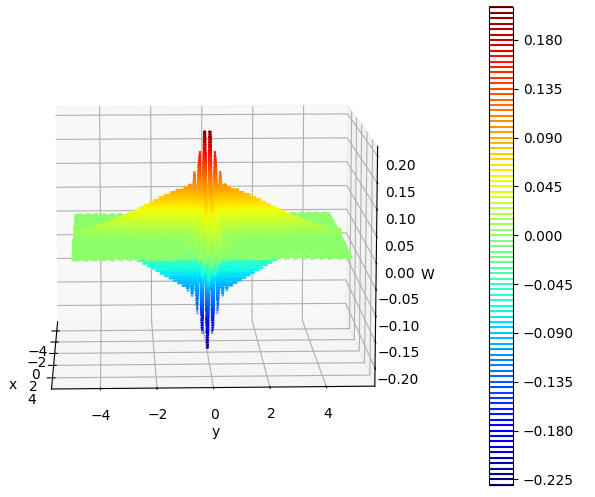}
                \caption{}
			\label{}
		\end{subfigure}
               \hfill
		\begin{subfigure}{0.4\textwidth}
			\includegraphics[width=1.2\linewidth]{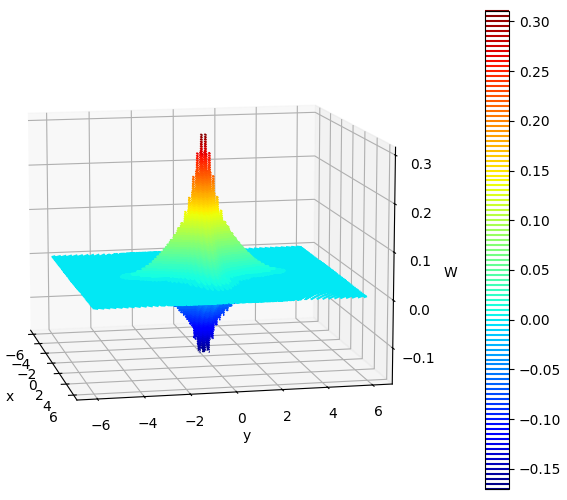}  
			\caption{}
                \label{}
		\end{subfigure}
		
\caption{The Wigner function for the ground state, obtained from the simulation of Eq. \ref{eq 2pRSM} for a $\frac{\gamma}{\omega}= 0.3$,  b $\frac{\gamma}{\omega}= 0.6$ and c $\frac{\gamma}{\omega}= 0.9$ . Here, we take the number of Fock states included in the cavity mode $N = 200$ in our numerical calculation.}
\label{W 2pRSM}
\end{figure}

In this section, we analyze the dynamics of the \textit{2pRSM} by studying the nonclassicality in the ground state of the system (Eq. \ref{eq 2pRSM}) under different values of the Stark coupling strength at the critical coupling strength $g$ and with number of cavity Fock states N = 200, by treating the negativity of its phase space distribution function such as Wigner function. 
To display this, we numerically calculate the Wigner function with a specific initial state $\rho_{GS}$ using QuTiP (Quantum Toolbox in Python) \cite{qutip}, as plotted in Fig. \ref{W 2pRSM}

For $\frac{\gamma}{\omega}= 0.3$, the Wigner function exhibits a symmetric, Gaussian-like distribution shape with sharp peaks concentrated around the origin and some oscillations in the phase space.\\
At this value of $\frac{\gamma}{\omega}$, the quantum state maintains a relatively coherent structure, with significant quantum interference indicated by the oscillations. This implies that the state operates within a regime where the quantum coherence remains comparably robust, and the Stark coupling has not appreciably compromised the coherence of the quantum state. As the Stark coupling strength gamma increases, the Wigner function still retains a symmetric form, however, the central peak is more localized, and the interference pattern has become more complex, this may indicate that the quantum state becomes more sensitive to the Stark coupling, leading to more significant quantum interference effects. The emergence of larger negative regions suggests an increase in non-classicality, potentially signifying the initiation of quantum squeezing or other non-classical features.\\
At a higher value of the Stark coupling strength ($\frac{\gamma}{\omega}= 0.9$), the Wigner function reveals a further enhancement in complexity, characterized by sharper peaks and more extensive negative regions, which indicates that the quantum state is likely in a more strongly coupled regime, wherein the Stark effect induces significant distortions in the quantum state. The increased non-classicality and complexity of the Wigner function suggest that the system is approaching a regime of spectral collapse or similar phenomena, in which the state’s coherence is significantly influenced by the Stark coupling.

\subsection{The Entanglement}
One of the remarkable phenomena associated with Rabi and realistic Stark-coupling term is the quantum phenomenon of entanglement, which is a fascinating feature of the quantum mechanics \cite{Einstein1935} and a cornerstone in various fields, such as quantum computation, quantum communication, quantum cryptography, quantum metrology, and quantum simulation.\\

The qubit-cavity field entanglement of any eigenstate $\ket{\Psi}$ of the \textit{2pRSM} can be quantified by the entanglement entropy which we choose here as the von Neumann entropy of the reduced density matrix for any one of the subsystems, e.g. the cavity field (cf) and qubit (q) subsystems considered here.

\begin{equation}
    S = -\Tr(\rho_{q} \log_{2} \rho_{q}) = -\Tr(\rho_{cf} \log_{2} \rho_{cf})
    \label{vn entropy}
\end{equation}

where $\rho_{q} = Tr_{cf} (\ket{\Psi}\bra{\Psi})$ and $\rho_{cf} = Tr_{q}(\ket{\Psi}\bra{\Psi})$.\\
The entanglement entropy vanishes for a separable state and equals one for a maximally entangled state.
We numerically evaluate the entropy expression \ref{vn entropy} by using QuTiP, and obtain the following result.\\

\begin{figure}[!h]
    \centering
    \includegraphics[width=0.8\linewidth]{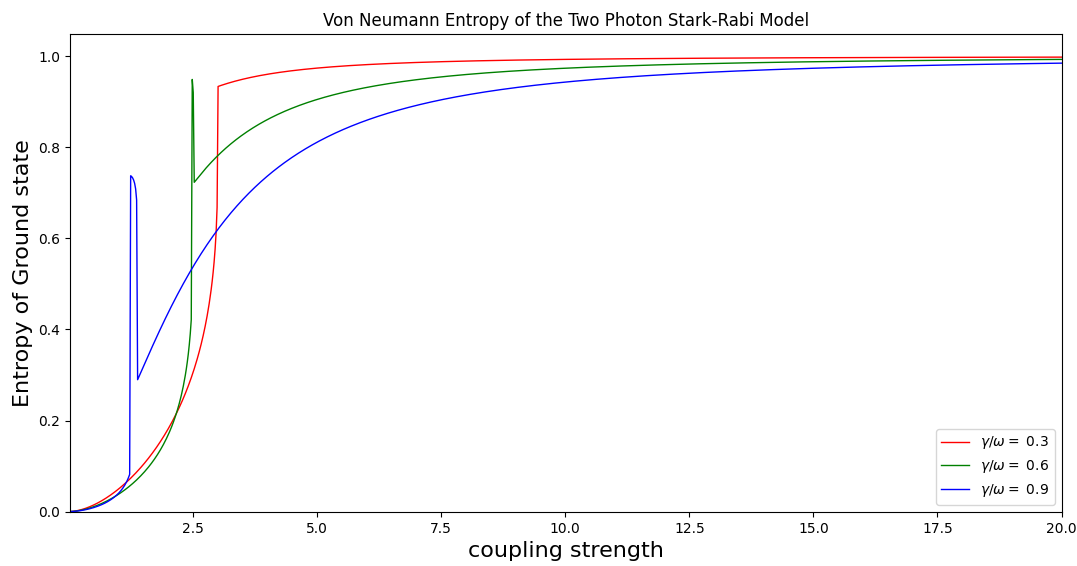}  
    \caption{von Neumann entropy S of the \textit{2pRSM} as a function of the coupling strength $g$ for various values of the Stark coupling strength $\frac{\gamma}{\omega}$. In particular, we use the following values: $\frac{\gamma}{\omega}= 0.3$ in red line, $\frac{\gamma}{\omega}= 0.6$ in green line, $\frac{\gamma}{\omega}= 0.9$ in blue line. }
    \label{Vn 2pRSM}
\end{figure}

In Fig \ref{Vn 2pRSM}, we plot the Von Neumann entropy of the ground state in the two-photon Rabi Stark model as a function of coupling strength. The different curves correspond to different ratios of the Stark coupling strength $\frac{\gamma}{\omega}$ show how the quantum entanglement changes under the effect of Stark coupling.
At low coupling strengths $g$, the entropy (and thus the entanglement) increases gradually. This indicates that the interaction between subsystems starts creating quantum correlations.\\
Where at the critical points of each curve the entropy increases rapidly, suggesting a phase transition or significant change in the system's quantum state. These points likely correspond to resonance conditions or significant changes in the system's dynamics.
Finally, the entropy tends to saturate, reaching a maximum value at high values of the coupling strength. This suggests that the system reaches a maximum entangled state, beyond which increasing the coupling strength doesn't significantly change the entanglement.\\
According to Figure \ref{Vn 2pRSM}, the quantum entanglement in the two-photon Stark-Rabi model is determined by both the coupling strength $g$ and the Stark coupling strength $\frac{\gamma}{\omega}$. These parameters allow one to influence the behavior of the system, including phase transitions.

\section{Conclusion}
\label{conclusion}

In this work, we have studied the quantum entanglement of the \textit{2pRSM}. As a first step, we provided a numerical simulation of the energy spectrum as a function the coupling strength $g$, at three different Stark coupling strength values $\frac{\gamma}{\omega}$. Also, we have used the Wigner function to investigate the phase space distribution in the ground state of the system as a function of the Stark coupling strength in correspondence of the critical coupling strength of each $\frac{\gamma}{\omega}$ value. Then, we have investigated the quantum entanglement of the system in the ground state as a function of the coupling in the same conditions.\\

To comprehensively explore the dynamics of the \textit{2pRSM}, we employ three approaches: numerical solution of the proposed model, the interpretation of the Wigner function in the ground state of the system, and the analysis of the system's dynamics through the von Neumann entropy.
We observed that, As the coupling strength increases, the entropy tends to augment, indicating enhanced entanglement between the field and the two-level system. At lower $\frac{\gamma}{\omega}$, the entanglement rises gradually, reaching a maximum value. However, at higher $\frac{\gamma}{\omega}$, the entropy curves exhibit sharp changes, including dips and jumps, which are likely associated with critical points and spectral collapse phenomena. These features suggest that the interplay between the Stark effect and coupling strength $g$ may leads to complex dynamics and phase transitions, profoundly influencing the entanglement properties of the system. These findings may have implications for cavity-based quantum computations.

\end{document}